\documentclass[a4paper,12pt]{article}
\usepackage{epsfig}
\usepackage{amssymb}
\usepackage{axodraw}

\newcommand{\beq}{\begin{equation}}
\newcommand{\eeq}{\end{equation}}
\newcommand{\bea}{\begin{eqnarray}}
\newcommand{\eea}{\end{eqnarray}}
\newcommand{\ba}{\begin{array}}
\newcommand{\ea}{\end{array}}
\newcommand{\bi}{\begin{itemize}}
\newcommand{\ei}{\end{itemize}}
\newcommand{\bn}{\begin{enumerate}}
\newcommand{\en}{\end{enumerate}}
\newcommand{\bc}{\begin{center}}
\newcommand{\ec}{\end{center}}

\newcommand{\no}{\nonumber}

\newcommand{\eq}[1]{eq.(\ref{#1})}

\newcommand{\gsim}{\lower.7ex\hbox{$\;\stackrel{\textstyle>}{\sim}\;$}}
\newcommand{\lsim}{\lower.7ex\hbox{$\;\stackrel{\textstyle<}{\sim}\;$}}

\begin{document}
\tolerance=100000
\thispagestyle{empty}
\setcounter{page}{0}

\begin{flushright}
{\rm CERN-TH/2003-183} \\
{\rm INFNNA-IV-03-30} \\
{\tt hep-ph/0308031}
\end{flushright}

\vspace*{\fill}

\begin{center}
{\Large \bf 
Soft Leptogenesis
}\\[2.cm]

{{\large\bf Giancarlo D'Ambrosio}$^1$,
{\large\bf Gian F. Giudice}$^2$
{\large and} {\large\bf Martti Raidal}$^{3}$}  
\\[7mm]
{\it $^1$ INFN, Sezione di Napoli and Dipartimento di Scienze Fisiche, \\ 
           Universit\`a di Napoli, I-80126 Napoli, Italy} \\[3mm]
{\it $^2$ Theoretical Physics Division, CERN, CH-1211 Geneva 23, Switzerland
} \\[3mm]

{\it $^3$ National Institute of Chemical Physics and Biophysics, 
Tallinn 10143, Estonia
} \\[10mm]
\end{center}

\vspace*{\fill}

\begin{abstract}
{\small\noindent

We study ``soft leptogenesis'', a new mechanism of leptogenesis 
which does not require flavour
mixing among the right-handed neutrinos. Supersymmetry soft-breaking terms
give a small mass splitting between the CP-even and CP-odd right-handed
sneutrino states of a single generation and provide a CP-violating phase
sufficient to generate a lepton asymmetry. The mechanism is successful
if the lepton-violating soft bilinear coupling is unconventionally
(but not unnaturally) small. The values of the right-handed neutrino masses
predicted by soft leptogenesis can be low enough to evade the cosmological
gravitino problem. 
}
\end{abstract}

\vspace*{\fill}

\begin{flushleft}
{\rm CERN-TH/2003-183}\\
{\rm  August 2003} \\
\end{flushleft}

\newpage
\setcounter{page}{1}

\section{Introduction}

After the experimental confirmation of neutrino oscillations,
leptogenesis~\cite{fy} has become the most economical and attractive
scenario to explain the cosmic baryon asymmetry. Within a range of neutrino
mass and mixing parameters compatible with experimental data, it successfully
reproduces the value $n_B/s=(0.87\pm 0.04)\times 10^{-10}$ 
derived from nucleosynthesis 
and CMB measurements. The see-saw mechanism~\cite{seesaw}
 employed in leptogenesis requires
the existence of right-handed neutrinos with masses close to the GUT scale.
Since both the stability of the GUT mass hierarchy and 
gauge coupling unification
strongly suggest low-energy supersymmetry, leptogenesis is more natural in a
supersymmetric framework. Once supersymmetry is introduced, sneutrino decays
offer a new channel for generating an asymmetry. 

In this paper we want to
discuss how the sneutrino decay channel is fundamentally different than the
neutrino channel. Supersymmetry-breaking terms remove the mass degeneracy
between the two real sneutrino states belonging to the supermultiplet
of a single neutrino generation~\cite{grossman}. They also provide a source
of CP violation, and 
the mixing between the two sneutrino states 
can generate a CP asymmetry in the decay.
Although the scale of supersymmetry-breaking is much smaller than
the right-handed neutrino mass, the asymmetry can be sizable because of the
resonant effect~\cite{resonant,p} 
of the two nearly-degenerate states. Contrary to leptogenesis
from neutrino decay, where at least two generations of right-handed
neutrinos are required, a 
single-generation right-handed sneutrino decay is sufficient to 
generate the CP asymmetry.
The soft terms, and not flavour physics, 
provide the necessary mass splitting and 
CP-violating phase. This new mechanism of leptogenesis, which
we will call ``soft leptogenesis'' can then be an 
alternative or an addition to the traditional scenario of mixing between 
different flavour states.

This paper is organized as follows. In sect.~2 we describe the one-generation
see-saw model in presence of supersymmetry-breaking effects and compute
the relevant CP asymmetry. In sect.~3 we rederive the asymmetry following
a different field-theoretical approach, and comment on the effect of the
initial-state coherence. The baryon-asymmetry efficiency factor 
is computed in sect.~4 by integrating the complete Boltzmann equations.
Finally our results for the baryon asymmetry are presented and discussed
in sect.~5.

As we were completing this work, a paper has appeared~\cite{nir}  
presenting the same idea.

\section{The CP Asymmetry}

The supersymmetric see-saw model is described by the superpotential
\beq
W=Y_{ij}N_iL_jH+\frac{1}{2}M_{ij}N_iN_j ,
\eeq
where $i,j=1,2,3$ are flavour indices and $N_i$, $L_i$, $H$ are the chiral 
superfields for the right-handed
neutrinos, the left-handed lepton doublets and the Higgs, respectively.
The supersymmetry-breaking terms involving the right-handed sneutrinos 
${\tilde N}_i$ are
\beq
-{\cal L}_{soft} = {\tilde m}^2_{ij}{\tilde N}_i^\dagger{\tilde N}_j
+\left(  A_{ij}Y_{ij}{\tilde N}_i {\tilde \ell}_jH+\frac{1}{2}B_{ij}
M_{ij}{\tilde N}_i{\tilde N}_j + {\rm h.c.}\right) ,
\eeq
with standard notations.

We will consider a single generation of $N$ because,
as explained in the introduction, our effect survives even in this limiting
case. For simplicity, we will also assume proportionality of soft trilinear
terms and drop the flavour index for the coefficient $A$. Under these
conditions, a  
CP-violating phase is still present. Indeed, with a superfield rotation
we can eliminate all phases from the superpotential parameters $Y_{1i}$
and $M$ ($\equiv M_{11}$), and with an $R$-rotation we can eliminate the
relative phase between $A$ and $B$. However, the remaining phase is physical.

The right-handed neutrino $N$ has a mass $M$, while sneutrino and antisneutrino
states mix in the mass matrix. Their mass eigenvectors
\bea
{\tilde N}_+ &=& \frac{1}{\sqrt{2}}\left( e^{i\Phi/2}{\tilde N}+
e^{-i\Phi/2}{\tilde N}^\dagger \right) \nonumber \\
{\tilde N}_- &=& \frac{-i}{\sqrt{2}}\left( e^{i\Phi/2}{\tilde N}-
e^{-i\Phi/2}{\tilde N}^\dagger \right) ,
\eea
with $\Phi \equiv {\rm arg}(BM)$, have mass eigenvalues
\beq
M_{\pm}^2=M^2+{\tilde m}^2\pm \left| BM \right| .
\eeq

The sneutrino interaction Lagrangian in the basis of 
flavour $({\tilde N},{\tilde N}^\dagger )$
and mass $({\tilde N}_+,{\tilde N}_-)$
eigenstates 
is, respectively,
 \bea
-{\cal L}_{int}&=&{\tilde N} \left( Y_{1i}{\bar{\tilde H}} \ell_L^i
+MY^*_{1i}{\tilde \ell}_i^* H^* +AY_{1i}{\tilde \ell}_iH \right) +{\rm h.c.}
\label{intlag}\\
&=& \frac{Y_{1i}}{\sqrt{2}}{\tilde N}_+ \left[{\bar{\tilde H}} \ell_L^i
+(A+M) {\tilde \ell}_iH \right] +i\frac{Y_{1i}}{\sqrt{2}}{\tilde N}_- 
\left[{\bar{\tilde H}} \ell_L^i
+(A-M) {\tilde \ell}_iH \right] +{\rm h.c.}
\label{intpiu}
\eea
Here, for simplicity, we have set $\Phi=0$ choosing, from now on, 
a basis where $A$ is
the only complex parameter.

The system of ${\tilde N}$--${\tilde N}^\dagger$ is completely analogous to
the $K^0$--${\bar K}^0$ or $B^0$--${\bar B}^0$ system, and in this section
we will treat it
with the same formalism (see {\it e.g.} ref.~\cite{nir2}). 
Its evolution is determined (in the
non-relativistic limit) by
the Hamiltonian $H={\hat M}-i{\hat \Gamma}/2$ where, at leading order in
the soft terms,
\bea
{\hat M}&=&M\pmatrix{1&\frac{\displaystyle{B}}{\displaystyle{2M}}\cr \frac{\displaystyle{B}}
{\displaystyle{2M}}&1},\\
\no \\
{\hat \Gamma}&=&\Gamma\pmatrix{1&\frac{\displaystyle{A^*}}{\displaystyle{M}}\cr \frac{\displaystyle{A}}
{\displaystyle{M}}&1}.
\eea
Here $\Gamma$ is the total $\tilde N$ decay width
\beq
\Gamma = \frac{(YY^\dagger )_{11}}{4\pi}M\equiv \frac{G_F}{\sqrt{2}\pi}
m M^2.
\eeq
With this (standard) definition, $m=(YY^\dagger)_{11}\langle H\rangle^2/M$ 
sets the scale for the physical (mainly
left-handed) neutrino masses $m_\nu^i$, 
since $m=\sum_i |r_i|^2m_\nu^i$, under the condition $\sum_i r_i^2=1$.

The eigenvectors of the Hamiltonian $H$ are
\bea
{\tilde N}_L &=& p{\tilde N}+q {\tilde N}^\dagger  \nonumber \\
{\tilde N}_H &=& p{\tilde N}-q {\tilde N}^\dagger ,  
\eea
\beq
\left( \frac{q}{p}\right)^2=\frac{{\hat M}_{12}^*-\frac{i}{2}{\hat
\Gamma}_{12}^*}{{\hat M}_{12}-\frac{i}{2}{\hat
\Gamma}_{12}} .
\label{pq}
\eeq

We consider an initial state at $t=0$ with equal densities of $\tilde N$
and ${\tilde N}^\dagger$. At time $t$, the state has evolved into
\bea
{\tilde N}(t) &=& g_+(t){\tilde N}(0)+\frac{q}{p} g_-(t){\tilde N}^\dagger(0)
  \nonumber \\
{\tilde N}^\dagger(t) &=&\frac{p}{q} g_-(t) {\tilde N}(0)+g_+(t)
{\tilde N}^\dagger(0) ~,
\label{timed}
\eea
\bea
g_+(t)&=&e^{-iMt}e^{-\Gamma t/2}\cos\left( \Delta M t/2\right) \nonumber \\
g_-(t)&=&ie^{-iMt}e^{-\Gamma t/2}\sin\left( \Delta M t/2\right) .
\eea
Here $\Delta M\equiv M_+-M_-=|B|$ and we have neglected $\Delta \Gamma$
with respect to $\Delta M$.

We can now compute the total integrated lepton asymmetry, defined by
\beq
\epsilon =\frac{\sum_f \int _0 ^\infty dt \left[ \Gamma ({\tilde N}(t)\to f)
+\Gamma ({\tilde N}(t)^\dagger\to f)- \Gamma ({\tilde N}(t)\to {\bar f})
-\Gamma ({\tilde N}(t)^\dagger\to {\bar f})\right]}
{\sum_f \int _{0} ^\infty dt \left[ \Gamma ({\tilde N}(t)\to f)
+\Gamma ({\tilde N}(t)^\dagger\to f)+ \Gamma ({\tilde N}(t)\to {\bar f})
+\Gamma ({\tilde N}(t)^\dagger\to {\bar f})\right]}.
 \eeq
Here $f$ is a final state with lepton number equal to 1 and $\bar f$ is
its conjugate. Since we want to exploit the enhancement due to the 
resonance~\cite{resonant,p}, 
we will disregard any other subleading effects. In particular,
we will neglect direct CP violation in the decay (vertex diagrams) and
include only the effect of the ${\tilde N}$--${\tilde N}^\dagger$ mixing
(wave-function diagrams). This means that the decay amplitudes of the
flavour sneutrino eigenstates can be immediately derived from the
interaction Lagrangian in eq.~(\ref{intlag}), setting $A=0$. We will include
the factors $c_F$ and $c_B$ to parametrize the phase space of the
fermionic ($f={\tilde H}\ell$) and bosonic ($f=H{\tilde \ell}$) final
states. Taking into account the time dependence described by eq.~(\ref{timed}),
the CP asymmetry is given by
\beq
\epsilon = \frac{1}{2} \left( \left| \frac{q}{p}\right|^2- 
\left| \frac{p}{q}\right|^2\right) \left( \frac{c_B-c_F}{c_F+c_B}\right)
\frac{\int_0^\infty dt \left| g_- \right|^2}
{\int_0^\infty dt \left( \left| g_+ \right|^2+\left| g_- \right|^2\right)}.
\eeq
Evaluating eq.~(\ref{pq}) in the limit ${\hat \Gamma}_{12}\ll
{\hat M}_{12}$, we find
\beq
\left|\frac{q}{p}\right|^2\simeq 1-{\rm Im}\frac{{\hat \Gamma}_{12}}
{{\hat M}_{12}}=1+\frac{2\Gamma ~{\rm Im} A}{BM}.
\eeq
Performing the time integral
\beq
\frac{\int_0^\infty dt \left| g_- \right|^2}
{\int_0^\infty dt \left( \left| g_+ \right|^2\left| +g_- \right|^2\right)}
=\frac{\Delta M^2}{2\left( \Gamma^2+\Delta M^2\right)},
\eeq
we obtain the final expression for the CP asymmetry
\bea
\epsilon &=& \frac{\Gamma B}{\Gamma^2+B^2}\frac{{\rm Im} A}{M}\Delta_{BF},
\label{epsilon}\\
 \Delta_{BF}&=&\frac{c_B-c_F}{c_F+c_B}.
\label{DFB}
\eea

It is easy to understand the origin of the different terms present 
in eq.~(\ref{epsilon}). The factor $AB$ signals the presence of
supersymmetry breaking and the violation of lepton number;
$(B/M){\rm Im}A$ signals CP violation. The resonance effect is described
by $\Gamma B /(\Gamma^2 +B^2)$, which is maximal when $\Gamma \sim |B|$. As we
move away from the resonance condition, 
$\epsilon$
suffers an extra power suppression. 

\begin{figure}[t]
\centerline{\epsfxsize = 0.5\textwidth \epsffile{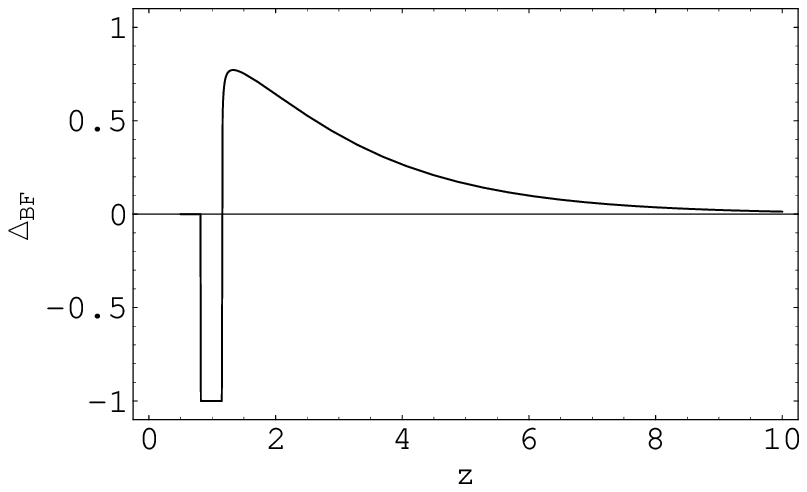}
%\hfill \epsfxsize = 0.5\textwidth \epsffile{}
}
\caption{\it
$\Delta_{BF}$, defined in \eq{DFB}, as a function of $z=M/T.$
\vspace*{0.5cm}}
\label{cfcb}
\end{figure}

An exact cancellation occurs between the asymmetry in the fermionic
and bosonic channels, if $c_F=c_B$. Thermal effects, which break supersymmetry,
remove this degeneracy. This happens both because of final-state Fermi 
blocking and Bose stimulation~\cite{roulet}, and because of the effective 
masses acquired by particle excitations inside the plasma (for a full
discussion of the thermal effects in leptogenesis, see ref.~\cite{noi}).
We find
\bea
c_F&=&(1-x_\ell -x_{\tilde H})\lambda(1,x_\ell,x_{\tilde H})
\left[ 1-n_F(E_\ell)\right] \left[ 1-n_F(E_{\tilde H})\right] 
\label{cfeq}\\
c_B&=&\lambda(1,x_H,x_{\tilde \ell})
\left[ 1+n_B(E_H)\right] \left[ 1+n_B(E_{\tilde \ell})\right]
\label{cbeq}
\eea
\bea
&E_{\ell ,{\tilde H}}=\frac{M}{2} (1+x_{\ell ,{\tilde H}}-
x_{{\tilde H},\ell}), ~~~
E_{H ,{\tilde \ell}}=\frac{M}{2} (1+x_{H ,{\tilde \ell}}-
x_{{\tilde \ell},H})&\\
&\lambda(1,x,y)=\sqrt{(1+x-y)^2-4x},~~~
x_a\equiv \frac{m_a(T)^2}{M^2}&\\
&n_F(E)=\frac{1}{e^{E/T}-1},~~~n_B(E)=\frac{1}{e^{E/T}+1},&
\eea
where the thermal masses for the relevant supersymmetric degrees of
freedom are
\bea
m_H^2(T)=2 m_{\tilde H}^2(T)&=& \frac{3}{8}g_2^2+\frac{1}{8}g_Y^2
+\frac{3}{4}\lambda_t^2,\\
m_{\tilde \ell}^2(T)=2 m_\ell^2(T)&=& \frac{3}{8}g_2^2+\frac{1}{8}g_Y^2.
\eea
Here $g_2$ and $g_Y$ are gauge couplings and $\lambda_t$ is the top Yukawa,
renormalized at the appropriate high-energy scale. The value of $\Delta_{BF}$
as a function of $z=M/T$ is plotted in fig.~\ref{cfcb}. Because of Bose
stimulation, $\Delta_{BF}$ is positive and grows with temperature. However,
for $z<1.2$, the sum of Higgs and slepton thermal masses
becomes larger than $M$, and the bosonic channel is kinematically closed.
Eventually, for $z<0.8$, also the fermionic channel becomes unaccessible.
This explains the abrupt changes of $\Delta_{BF}$ shown in fig.~\ref{cfcb}.

\section{Field-Theoretical Approach}

\begin{figure}[t]
\begin{center}
\begin{picture}(320,80)(0,0)
\ArrowLine(130,0)(90,40)
\ArrowLine(90,40)(130,80)
\DashLine(20,40)(90,40){3}
\Text(55,50)[]{\small $\tilde N_-$}
\Text(135,70)[]{\small $\ell$}
\Text(135,13)[]{\small $\tilde H$}
%\Text(110,40)[]{\small $f_-$}
%%%%%%%%%%
\ArrowLine(320,0)(280,40)
\ArrowLine(280,40)(320,80)
\DashLine(180,40)(280,40){3}
\Vertex(230,40){8}
\Text(205,50)[]{\small $\tilde N_-$ }
\Text(265,50)[]{\small $\tilde N_+$ }
\Text(325,70)[]{\small $\ell$}
\Text(325,13)[]{\small $\tilde H$}
%\Text(300,40)[]{\small $f_+$}
%%%%%%%%%%%%%%%%%%%%%%
\end{picture}
\end{center}
\caption{\it 
Interfering $\tilde N_-$ decay amplitudes for the fermionic final states. 
Analogous
diagrams exist for bosonic final states. 
The two-point function $\Pi_{+-}$ denoted by a blob
contains a sum of all  possible intermediate states.
}  
\label{fig1}
\end{figure}
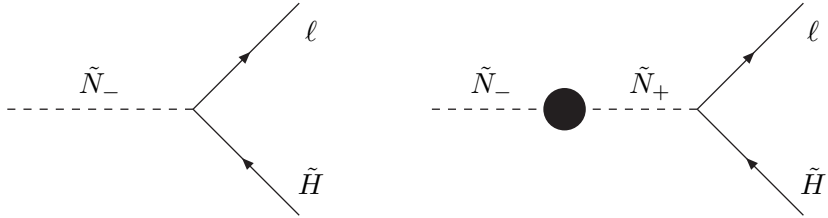

In this section we want to study the  CP asymmetry using a different
procedure. 
We use an effective field-theory approach 
 of resummed propagators 
for unstable (mass eigenstate) particles, as described in ref.~\cite{p}.
 The decay
amplitude ${\widehat f}_-$ 
of the unstable external state ${\tilde N}_-$ into a final state
$f$ is described by a superposition of amplitudes with stable external
states $f_\pm$. Adding 
the contributions shown in fig.~\ref{fig1}, we obtain
\beq
\widehat{f}_-(\tilde N_-  \rightarrow f ) = f_- - f_{+} 
\frac{i\Pi _{+-}}{M_-^2 -M_+^2 + i\Pi _{++}} ,
\label{ampl}
\eeq
where $\Pi _{ab} (p^2)$ are the absorptive parts of the
two-point functions for $a,b=+,-,$ which,
in our case, are given by 
\beq
\Pi_{++}=\Pi_{--}=M\Gamma,~~~\Pi_{+-}=\Pi_{-+}=- {\rm Im}A\Gamma .
\eeq
The amplitude for the decay into the conjugate final state is
\beq
\widehat{\bar f}_-(\tilde N_-  \rightarrow {\bar f} ) = f_-^* - f_{+}^* 
\frac{i\Pi _{+-}}{M_-^2 -M_+^2 + i\Pi _{++}} .
\eeq

Squaring the amplitudes and multiplying by the
phase-space factors $c_F$ and $c_B$, we obtain the asymmetry
\bea
\epsilon_-&=&\frac{\sum_f \left[\Gamma(\tilde N_- \rightarrow f)-
\Gamma(\tilde N_- \rightarrow {\bar f})\right]}
{\sum_f \left[\Gamma(\tilde N_- \rightarrow f)+
\Gamma(\tilde N_-  \rightarrow {\bar f})\right]}
\\ &= &
\frac{2  \left( M_-^2 - M_+^2 \right) \,
\sum_f {\rm Im}\left( f_-^* f_+\right) \Pi_{+-} c_f}
{\sum_f \left[ |f_-|^2\left( M_-^2 - M_+^2 \right)^2 +
\left|f_- \Pi_{++} - f_+ \Pi_{+-}) \right|^2\right] c_f}.
\label{eq:epsg}
\eea
The corresponding results for $\tilde N_+$ are obtained by interchanging 
the indices $+$ and $-$. 

Neglecting supersymmetry-breaking 
in vertices, from the interaction Lagrangian in eq.~(\ref{intpiu})
we obtain, up to an overall normalization, $f_+=1$, $f_-=-i$ for
the scalar-channel final state (Higgs and slepton) and
$f_+=1$, $f_-=i$ for the fermionic channel (higgsino and lepton).
Inserting these values in eq.~(\ref{eq:epsg}) and combining the
asymmetries from ${\tilde N}_-$ and ${\tilde N}_+$, we obtain the
final expression for the total CP asymmetry
\beq
\epsilon = \frac{4\Gamma B }{4 B ^2 +\Gamma ^2}\frac{{\rm Im} A}{M}\Delta_{BF}.
\label{epsilon2}
\eeq

This result agrees with eq.~(\ref{epsilon}) in the limit $\Gamma \ll \Delta
M$. When $\Gamma \gg \Delta M$, the two states are not well-separated 
particles. Therefore, the result for the asymmetry depends on how the
initial state is preparated. If sneutrinos (like $K$ and $B$) are
produced in current eigenstates and evolve freely ({\it e.g.} if produced
in inflaton decay out of equilibrium), the formalism followed in sect.~2
gives the correct answer, taking into account the coherence of the initial
state. On the other hand, if $\tilde N$ are in a thermal bath with a
thermalization time $\Gamma^{-1}$ shorter than the oscillation time 
$\Delta M^{-1}$, coherence is lost and eq.~(\ref{epsilon2}) gives a more
appropriate description.
Therefore in principle we are sensitive to the details of
the initial state.
In practice, the difference is inessential since we can just recast 
eq.~(\ref{epsilon}) into eq.~(\ref{epsilon2}) with a redefinition of
the unknown soft parameters, $A\to 2A$, $B\to 2B$. In the following,
we will use eq.~(\ref{epsilon2}) in our discussion.

%\begin{figure}[htbp]

\section{Solutions to the Boltzmann Equations}

The baryon asymmetry is given by 
\beq
\frac{n_B}{s}=-\left( \frac{24+4n_H}{66+13n_H}\right)
\frac{\epsilon}{\Delta_{BF}} ~\eta ~Y^{eq}_{\tilde N}.
\eeq
The first factor~\cite{tur} takes into account the reprocessing of the $B-L$
asymmetry by sphaleron transitions, with the number of Higgs doublets $n_H$
equal to 2. $Y^{eq}_{\tilde N}=45\zeta (3)/(\pi^4 g_*)$ is
the sneutrino equilibrium density in units of entropy density, for 
temperatures much larger than $M$. For the minimal supersymmetric model
with one generation of right-handed neutrinos, the number of effective
degrees of freedom is $g_*=225$. Then, we obtain
\beq
\frac{n_B}{s}=-8.6\times 10^{-4}\frac{\epsilon}{\Delta_{BF}} ~\eta .
\label{baric}
\eeq

The efficiency factor $\eta$ describes the effects caused by: 
{\it i)} the
sneutrino density being smaller than the equilibrium density, {\it ii)} 
the wash-out from the lack of perfect out-of-equilibrium decay,
{\it iii)} the temperature-dependence of $\epsilon$ through $\Delta_{BF}$.
It is
obtained by integrating the relevant Boltzmann equations. We 
have  numerically solved
the set of differential equations describing decay, inverse decay, and
scattering processes for all supersymmetric particles, including thermal
masses for the particles involved~\cite{noi}.
With our definition of $\eta$, the temperature-dependent part $\Delta_{BF}$
has been factored out from $\epsilon$, see eq.~(\ref{baric}). We have included
in $\Delta_{BF}$ thermal masses and final-state statistical factors, as
described by eqs.~(\ref{cfeq})--(\ref{cbeq}), but we have neglected thermal
corrections to the loop diagram generating the asymmetry (for complete 
expressions of the thermal
corrections, see ref.~\cite{noi}).

\begin{figure}[t]
\centerline{\epsfxsize = 0.6\textwidth \epsffile{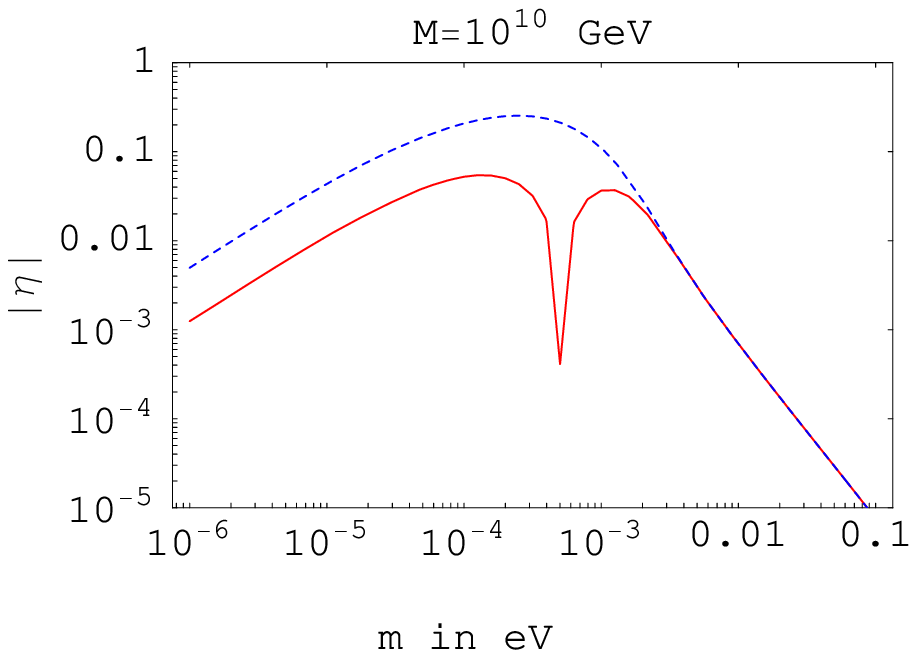}
\hfill \epsfxsize = 0.4\textwidth \epsffile{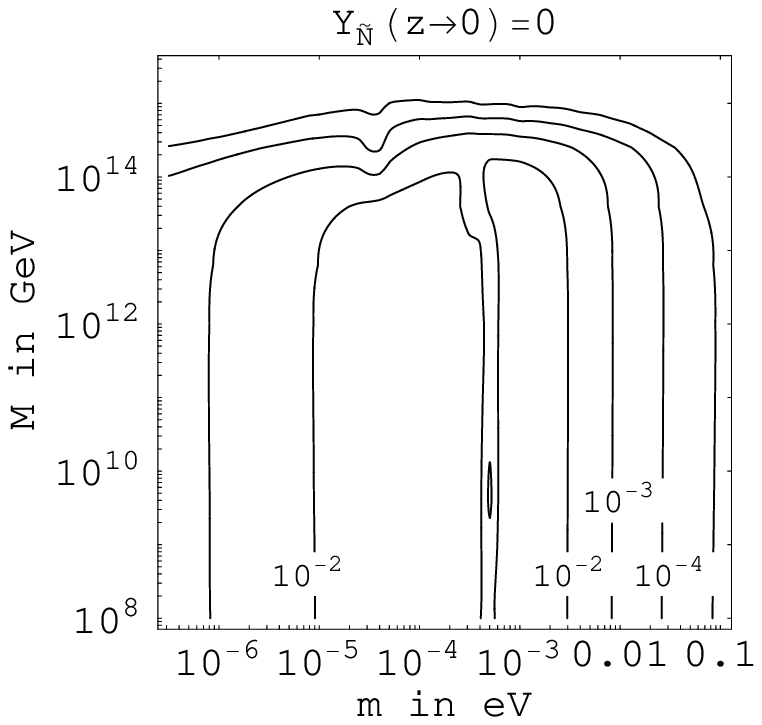}
}
\caption{\it
Left: Efficiency $|\eta |$ as a function of $m$ for $M=10^{10}$ GeV
and for two different
initial conditions: (i) vanishing initial $\tilde N$ abundance 
(solid red curve); (ii) thermal initial $\tilde N$ abundance, 
$Y_{\tilde N}^{eq}(z\to 0)$ (short-dashed blue curve).
 Right: isocurves of  $|\eta|=10^{i},$ $i=-2,-3,-4$ 
on the $(m,M)$ plane for the case (i).
\vspace*{0.5cm}}
\label{eff}
\end{figure}

In  
fig.~\ref{eff} (left) we plot the absolute value of the efficiency 
$\eta$ as a function of $m$ for fixed  $M=10^{10}$ GeV.
We consider
two different initial conditions for $Y_{\tilde N}$, the sneutrino density
in units of the entropy density.
In the first case, we assume that the ${\tilde N}$ population is created
by their Yukawa interactions with the thermal
plasma, and set $Y_{\tilde N}(z\to 0)=0$. The second case corresponds to
an initial $\tilde N$
abundance equal to the thermal one, 
$Y_{\tilde N}(z\to 0)= Y_{\tilde N}^{eq}(z\to 0)$. Here we are assuming that 
some unspecified high-energy interaction ({\it e.g.} GUT couplings)
is responsible for bringing the sneutrinos into an equilibrium density
at $T\gg M$.
In fig.~\ref{eff} (right) 
we present isocurves of  
$|\eta|=10^{i},$ $i=-2,-3,-4$ 
on the $(m,M)$ plane, for the initial condition
$Y_{\tilde N}(z\to 0)=0$. This demonstrates that
the efficiency is almost independent of $M$, in the 
range of $M$ that is relevant for us.

The results in Fig.~\ref{eff} indicate that, because of $\Delta_{BF}$, 
there is an
extra suppression of the soft-leptogenesis efficiency 
compared to the standard
leptogenesis case. Notice that this suppression occurs also if 
$Y_{\tilde N}(z\to 0)= Y_{\tilde N}^{eq}(z\to 0)$ (dashed line).
The smaller $m$, the stronger the suppression, because the out-of-equilibrium
decay occurs at lower $T$, where $\Delta_{BF}$ is smaller, see fig.~\ref{cfcb}.

\begin{figure}[t]
\centerline{\epsfxsize = 0.5\textwidth \epsffile{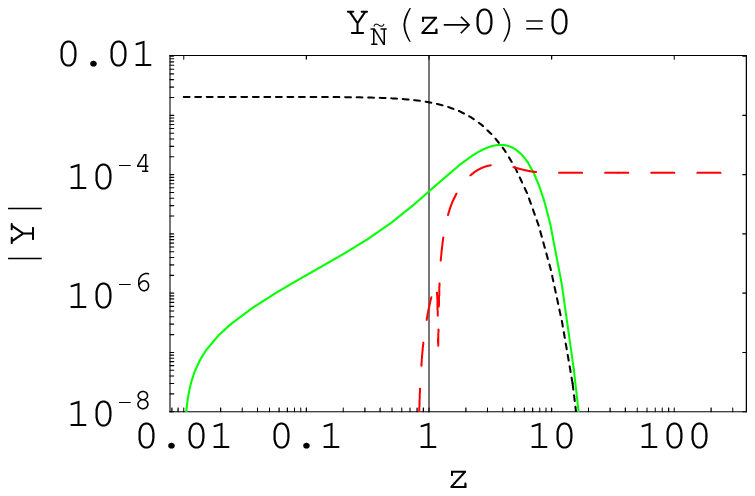}
\hfill \epsfxsize = 0.5\textwidth \epsffile{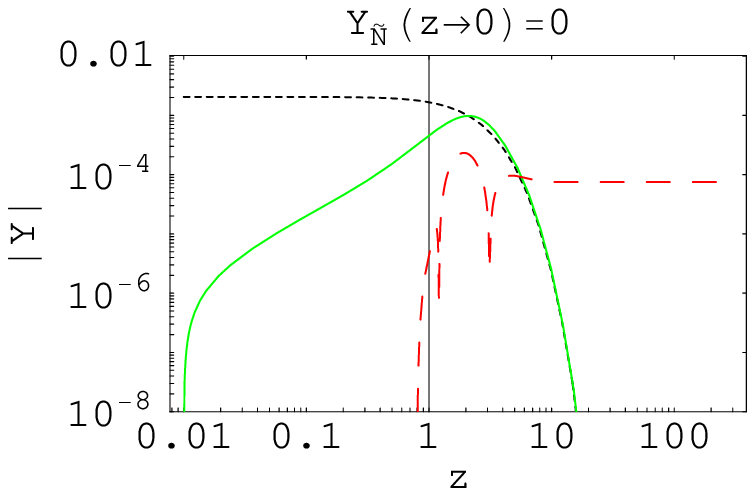}
}
\caption{\it
Evolution of the absolute values of the 
abundances $|Y_X|$ with $z=M/T$ for $M=10^{10}$ GeV, 
$m=10^{-4}$~eV (left) and $m=10^{-3}$~eV (right).
$Y_{\tilde N}^{eq}(z)$ is denoted with short-dashed black line, 
$Y_{\tilde N}(z)$
by the solid green line,  while the red long-dashed line
denotes the lepton asymmetry
$Y_L(z)/\epsilon_{const}$ with $\epsilon_{const}=10^{-6}$. 
\vspace*{0.5cm}}
\label{evol}
\end{figure}

In the case
$Y_{\tilde N}(z\to 0)=0$ (solid line) we observe 
a double-peak structure in $|\eta|$.
To understand this behaviour we plot in  Fig.~\ref{evol}
the evolution of the abundances with $z$ for 
$M=10^{10}$ GeV and $m=10^{-4}$~eV (left), $m=10^{-3}$~eV (right).
The solid green lines denote $Y_{\tilde N}(z)$ and 
the red long-dashed lines denote the lepton asymmetries 
$Y_L(z)/\epsilon_{const}$, for a fixed arbitrary value 
$\epsilon_{const}=10^{-6}.$
For reference, we also plot the equilibrium
density $Y_{\tilde N}^{eq}(z)$ with the short-dashed black line.

For $z<1.2$ the fermionic channel of sneutrino (inverse)
decay creates an asymmetry.
As soon as the bosonic channel is open ($z>1.2$, see fig.~\ref{cfcb}), 
it dominates and the asymmetry flips sign. This is illustrated by the
dip of the dashed lines at $z=1.2$ in 
fig.~\ref{evol} (both left and right), but this effect is inconsequential
for the final asymmetry. 
During the ${\tilde N}$-production phase, a {\it wrong}-sign asymmetry is 
generated compared to the  {\it right}-sign asymmetry  
produced in  ${\tilde N}$ decays.
For small $m$ (fig.~\ref{evol} left) the Yukawa interactions are weak and
the decay occurs at late time (small $T$) when $\Delta_{BF}$ is small.
Therefore
the generation of the {\it right}-sign asymmetry cannot overcome the
{\it wrong}-sign asymmetry. 
For larger $m$ (fig.~\ref{evol} right) the washout of the initial
 {\it wrong}-sign asymmetry is more efficient, and  at late time an 
asymmetry with the {\it right} sign is created. This is observed 
in the right plot of fig.~\ref{evol} as 
the additional sign-flip of $Y_L$ (or dip of the long-dashed curve).
At an intermediate value of $m$ the two effects perfectly
compensate each other, and the final asymmetry vanishes, as shown in 
fig.~\ref{eff} (solid line). 
In the case of an initial thermal ${\tilde N}$ distribution
(dashed line in fig.~\ref{eff}) this cancellation never occurs, since the
production phase is irrelevant.

\section{Discussion of the Results}

We now have all the ingredients to discuss the results of the baryon asymmetry
generated by the proposed mechanism of soft leptogenesis. 
The CP asymmetry is maximal
when the parameters lie on the resonance condition, $\Gamma =2|B|$, where
eq.~(\ref{epsilon2}) becomes
\beq
\frac{\epsilon}{\Delta_{BF}} =\frac{{\rm Im} A}{M}.
\eeq
From eq.~(\ref{baric}) 
and from the results shown in fig.~\ref{eff}, we obtain that
the presently observed baryon asymmetry requires\footnote{The
proportionality of the trilinear soft terms, assumed here, is certainly 
a questionable hypothesis and
should not be strictly applied. However,
the stability of the electroweak vacuum implies a bound on the size
of $A$. Let us consider the see-saw one-generation
model 
along a $D$-flat and $F_N$-flat ($\partial W/\partial N =0$) direction
$$
{\tilde \ell}=\pmatrix{0\cr \phi},~~~
H=\pmatrix{\phi \cr 0},~~~
{\tilde N}=-\frac{\lambda \phi^2}{M}.
$$
The scalar potential becomes
$$
V=\frac{2M}{Y^2}\left[ x^3+\left(\frac{B}{2}-A\right)x^2+{\tilde m}^2
x\left( 1+\frac{x}{2M}\right)\right] ,
\nonumber
$$
where $x\equiv Y^2 \phi^2/M$ and, for simplicity, we have taken equal soft
masses $\tilde m$ for all scalar fields. Minima  of the potential occur at
$x=[A-B/2\pm \sqrt{(A-B/2)^2-3{\tilde m}^2}]$. The request that the
potential is positive at these minima (to avoid instabilities of the 
electroweak vacuum) leads to the condition $|A-B/2|<2{\tilde m}$. This
shows that a departure from proportionality cannot significantly
enhance the CP asymmetry,
unless we accept to live on metastable vacua.}
\beq
M< \frac{{\rm Im} A}{\rm TeV}~  10^{8-9} ~{\rm GeV} .
\label{bounm}
\eeq

The resonance condition $\Gamma =2 |B|$ occurs when
\beq
M=\left( \frac{10^{-3}~{\rm eV}}{m}\right)^{1/2}
\left( \frac{B}{100~{\rm GeV}}\right)^{1/2}10^{10}~{\rm GeV}.
\label{mres}
\eeq
For typical values of $B$ around the electroweak scale, the value
of $M$ in eq.~(\ref{mres}) is larger than what is required by
eq.~(\ref{bounm}), and 
$n_B/s$ is predicted to be too small. Soft leptogenesis can
give a significant contribution to the baryon asymmetry only for very small
values of $B$.

\begin{figure}[t]
\centerline{\epsfxsize = 0.47\textwidth \epsffile{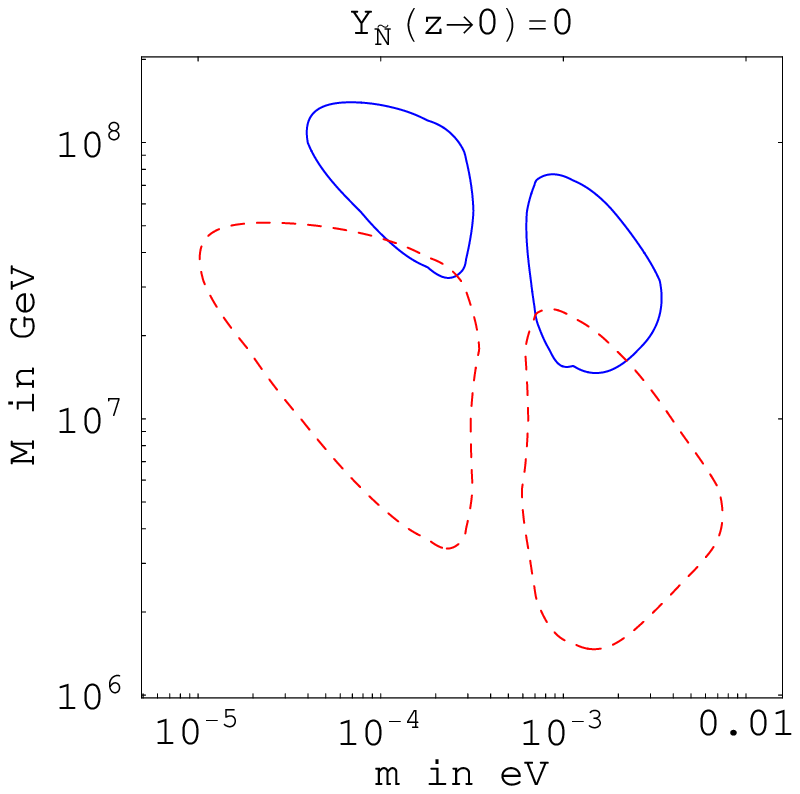}
\hfill \epsfxsize = 0.47\textwidth \epsffile{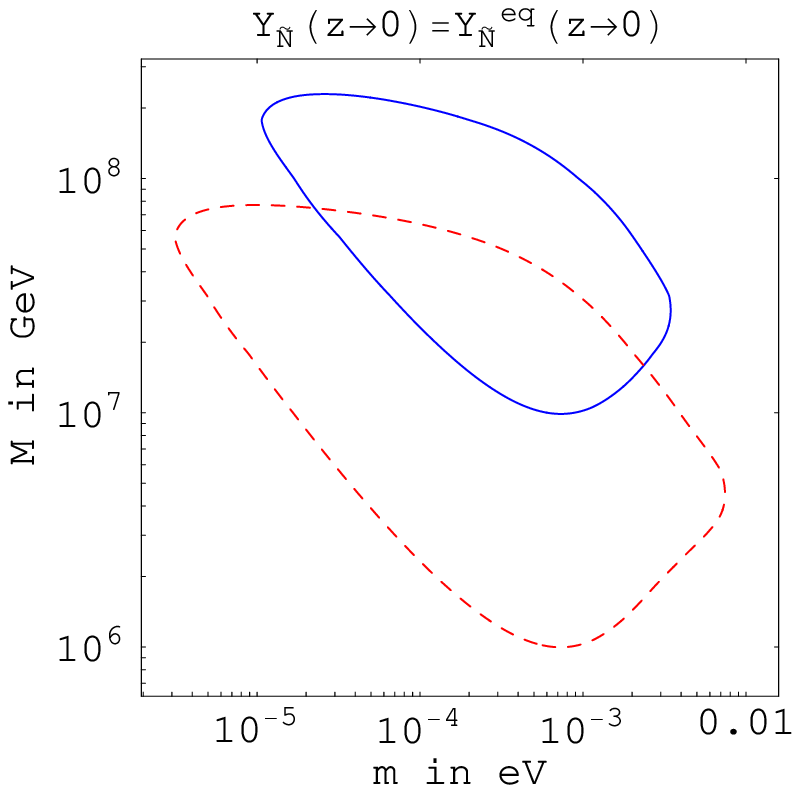}
}
\caption{\it
Regions of
$(m,M)$ plane where soft leptogenesis 
predicts $n_B/s>0.83\times 10^{-10}$ for ${\rm Im}A<{\rm TeV}$
and $B_M=100$~GeV (dashed line) and 1~TeV (solid line).
Soft leptogenesis is successful inside the contours.
We have assumed a vanishing 
(left) or thermal (right) initial
sneutrino density.
\vspace*{0.5cm}}
\label{mm}
\end{figure}

Very low values of $B$ require that 
the lepton-violating bilinear soft term
should not be generated at the leading order in supersymmetry breaking, 
but only by some 
higher-dimensional operators. Let us consider the supersymmetry-breaking
spurion superfield $X=\theta^2{\tilde m}M_{\rm Pl}$. Our assumption
is that the leading contribution to $B$, coming from the operator
$\int d^2\theta XMN^2/M_{\rm Pl}$, vanishes. In a general supergravity
scenario, this is not the case. One can however envisage dynamical
relaxation mechanisms (see {\it e.g.} ref.~\cite{jap}) which set $B=0$
at leading order.
Then, $B$ is determined by the operator in the K\"ahler potential
$\int d^4\theta XX^\dagger N^2/M_{\rm Pl}^2$, which gives a value 
$B\sim {\tilde m}^2/M$. The resonance condition $\Gamma =2|B|$ in terms of 
\beq
B_M\equiv \sqrt{BM},
\label{cafon}
\eeq
is
\beq
M=\left( \frac{10^{-3}~{\rm eV}}{m}\right)^{1/3}
\left( \frac{B_M}{100~{\rm GeV}}\right)^{2/3}2\times 10^{7}~{\rm GeV}.
\label{bifolc}
\eeq
According to our previous hypothesis, $B_M$ is the parameter to be taken of
the order of the electroweak scale. In this case, the value of $M$ in
eq.~(\ref{bifolc}) is in agreement with  eq.~(\ref{bounm}).

Our hypothesis of a small value of $B$ 
is  not technically unnatural. Indeed, radiative corrections
to the lepton-violating bilinear term are of the form $\delta B
\sim (YY^\dagger )_{11} A \ln (\Lambda^2/M^2)/(16 \pi^2) \sim 
\Gamma A/ M$, where $\Lambda$ is some ultraviolet cutoff scale. 
Thus, $\delta B$ is
much smaller than the assumed tree-level value ($B\sim \Gamma $). 
On the other hand, we stress that it would
have been unnatural to choose a very small trilinear coefficient, since $A$ 
receives gauge radiative corrections.

In fig.~\ref{mm} we quantify our results by showing the regions of parameters 
in the 
$(m,M)$ plane where soft leptogenesis 
can predict $n_B/s=(0.87\pm 0.04)\times 10^{-10}$ for ${\rm Im}A<{\rm TeV}$
and $B_M$ between 100~GeV (dashed line) and 1~TeV (solid line).
Soft leptogenesis is successful in the $(m,M)$ region inside the contours.
The two plots (left and right) correspond to vanishing and thermal initial
sneutrino density, respectively. There is no overlap between the region
of $(m,M)$ parameters favourable for soft leptogenesis with the one
suggested by conventional leptogenesis.

The values of $M$ required by soft leptogenesis (see fig.~\ref{mm}) are 
smaller
than the usual see-saw expectation, and imply very small Yukawa
couplings, $Y<10^{-4} (M/10^7~{\rm GeV})^{1/2}$. It should be said that
soft leptogenesis is more natural in presence of a large 
mass hierarchy of right-handed neutrinos,
since one is working in the one-generation limit.
Therefore it is not inconsistent to predict that one generation of $N$ 
lies at 
a mass scale significant lower than the GUT scale. 

This result has interesting consequences for the gravitino problem. In
traditional leptogenesis, the mass of the right-handed neutrino is bounded
from below~\cite{davids}, $M>2.4 (0.4)\times 10^9$~GeV for vanishing (thermal)
initial neutrino densities~\cite{altri}. Such values of $M$ 
are often uncomfortably large  when
compared with the upper bounds on the reheat temperature after inflation
$T_{RH}<10^{8-10}$, obtained by the 
requirement that relic gravitinos do not upset
the successful predictions of nucleosynthesis~\cite{sarkar}. 
On the other hand, soft leptogenesis
needs values of $M$ in the range $10^{6-8}$~GeV, well within the limits
imposed by the gravitino cosmological problem.

In conclusion, we have discussed how soft leptogenesis provides
an interesting interplay between lepton-number violating interactions at
high energy and low-energy supersymmetry-breaking terms. 
We have found that soft leptogenesis
can explain
the observed baryon asymmetry within the range of parameters shown
in fig.~\ref{mm}.
This requires {\it i)} an unconventional
(but not unnatural) choice of the lepton-violating bilinear soft parameter,
such that $B_M$ in eq.~(\ref{cafon})  of the order of the electroweak
scale; {\it ii)}
values of $M$ in the range $10^{6-8}$~GeV, which is favourable 
to evade the gravitino problem.

\section{Acknowledgements}

We thank R.~Rattazzi for useful discussions and the authors of ref.~\cite{nir}
for communicating with us.
The work of G.D. is supported in
part by TMR, EC--Contract No. ERBFMRX--CT980169 (EURODA$\Phi $NE)
and the one of M.R. by the ESF grant No. 5135.

\end{document}